\documentclass[conference]{IEEEtran}
\IEEEoverridecommandlockouts

\usepackage{cite}
\usepackage{amsmath,amssymb,amsfonts}
\usepackage{algorithmic}
\usepackage{graphicx}
\usepackage{xcolor}

\usepackage[utf8]{inputenc}
\usepackage{newtxtext,newtxmath}
\usepackage{caption}
\usepackage{subfig}
\usepackage{lineno,hyperref}
\modulolinenumbers[5]
\usepackage{comment}
\usepackage{multirow}
\usepackage{makecell}
\usepackage{tabularx}

\def\BibTeX{{\rm B\kern-.05em{\sc i\kern-.025em b}\kern-.08em
    T\kern-.1667em\lower.7ex\hbox{E}\kern-.125emX}}
\begin{document}

\title{PROFET: \textit{PROF}iling-based CNN Training Latency Proph\textit{ET} for GPU Cloud Instances}

\author{\IEEEauthorblockN{Sungjae Lee, Yoonseo Hur, Subin Park, Kyungyong Lee}
        \IEEEauthorblockA{Department of Computer Science, Kookmin University, Seoul, South Korea\\
                        \{sungjae, yoonseo, subean, leeky\}@kookmin.ac.kr
        }
}
\maketitle

\begin{abstract}
Training a Convolutional Neural Network (CNN) model typically requires significant computing power, and cloud computing resources are widely used as a training environment. However, it is difficult for CNN algorithm developers to keep up with system updates and apply them to their training environment due to quickly evolving cloud services. Thus, it is important for cloud computing service vendors to design and deliver an optimal training environment for various training tasks to lessen system operation management overhead of algorithm developers. To achieve the goal, we propose PROFET, which can predict the training latency of arbitrary CNN implementation on various Graphical Processing Unit (GPU) devices to develop a cost-effective and time-efficient training cloud environment. Different from the previous training latency prediction work, PROFET does not rely on the implementation details of the CNN architecture, and it is suitable for use in a public cloud environment. Thorough evaluations reveal the superior prediction accuracy of PROFET compared to the state-of-the-art related work, and the demonstration service presents the practicality of the proposed system.
\end{abstract}


\section{Introduction}\label{sec:introduction}
Deep Neural Network (DNN) has paved the way for the application of artificial intelligence algorithms in many fields, including image recognition and natural language processing. The success of deep learning is due to the advancement in DNN algorithms~\cite{deeplearning-neuralnet}, programming interfaces and platforms~\cite{pytorch-short, tensorflow-short, theano-short}, and optimized hardware for deep learning, such as the GPU and Tensor Processing Unit (TPU)~\cite{tpu}. 

Among many DNN algorithms, Convolutional Neural Network (CNN)~\cite{lenet-short, vgg, alexnet-short} is widely used for various applications, such as object detection for self-driving cars and image recognition. The complex internal structure of CNN implementations requires massive computing power and parallelism, particularly during the training step. Depending on the model complexity, training a CNN model can take a few weeks or months~\cite{cnn-detail, alibaba-dl-train-workload-short}, and building a cost-efficient and fast processing environment is crucial to increase the productivity of CNN algorithm developers. In addition, GPU devices can provide massive parallelism, but building a GPU cluster can be prohibitively expensive. Users can easily build a CNN model without purchasing GPU devices by using cloud resources. Training jobs may occur on occasion, and using GPU instances only when needed can result in cost savings.

The pace of cloud computing innovation is very fast, with new types of cloud services and instances being released on a regular basis by public vendors. It is difficult for deep learning algorithm developers to keep up with updates from public cloud vendors and understand the impact of the updates in order to apply them to their deep learning training pipeline. As a result, it is ideal if cloud service vendors provide an optimized deep learning environment natively so that algorithm developers can focus on core application development.

The Platform-as-a-Service (PaaS) cloud service model promoted as reducing development platform management overhead, allowing programmers to focus on critical software development tasks. Current deep learning pipeline platform services provided by public cloud vendors, such as AWS SageMaker and Google Cloud Datalab, are far from optimal because users must manually select cloud instance types and the number of instances to train a DNN model. Furthermore, an optimal DNN environment can vary across model architectures, dataset sizes, and model configurations, making deep learning platform management more difficult.

To understand the performance characteristics of the various CNN algorithm training latency on different public cloud GPU instances, we measured the training time on AWS EC2 instances that are equipped with GPU devices (Table~\ref{tab:aws_gpu_spec}), and the result is presented in Figure~\ref{fig:motivation_model}. Different CNN model architectures could result in a five-fold performance difference between the best and worst-performing GPU instance types. Despite such stark performance differences when training a CNN model on GPU cloud instances, algorithm developers rarely understand such performance characteristics and miss opportunities to optimize training environments. In addition, imposing training environment operation burdens on DNN algorithm developers might be too much overhead for them, as they are not generally system experts. Thus, the cloud service vendor should prepare an optimal training environment that reflects algorithm developers’ implementation characteristics.

Habitat~\cite{habitat-short}, Paleo~\cite{paleo-short}, NeuralPower~\cite{neuralpower-short}, and MLPredict~\cite{mlpredict-bigdata-short} proposed algorithms to predict the training time for various CNN algorithms on different GPU devices. The systems use the internal architecture of a CNN implementation and GPU device characteristics as input features. We call such approaches \emph{white-box} methods. Though they envisioned the feasibility of predicting the training time of CNN implementations on arbitrary GPU devices, the \emph{white-box} approach can be challenging to apply in an environment where a deep learning algorithm developer and a DNN development platform provider, such as a public cloud vendor, are not identical. In public cloud services, to provide an optimal DNN training environment using a \emph{white-box} approach, a service provider should know details of the model architecture. However, it is unlikely that algorithm developers would be willing to share the source code, which is generally confidential and is a private asset of an organization.

To overcome the shortcomings of the previous works, we propose PROFET, which aims to predict the training time for arbitrary CNN implementations without revealing the internal model architecture. Hiding the model architecture in a training latency prediction model makes the proposed system appropriate for a public cloud system where algorithm developers and development platform maintainers are different. To meet the goal, PROFET uses abstracted profiling information from CNN training as prediction model input and proposes novel heuristics of median-ensemble modeling to enhance prediction accuracy. PROFET can predict the CNN model training time on diverse GPU devices to support CNN training scenarios on the cloud GPU instances. A thorough evaluation of PROFET reveals that it outperforms the state-of-the-art algorithms, Habitat~\cite{habitat-short}, Paleo~\cite{paleo-short}, and MLPredict~\cite{mlpredict-bigdata-short}, improving prediction accuracy by $32\%$, $68\%$, and $82\%$, respectively. 

In summary, the major contributions are as follows.
\begin{itemize}
\item{Envisioning the importance of \emph{black-box} performance estimation of the CNN on a public cloud}
\item{Unique median-ensemble modeling to predict training latency across distinct GPU devices}
\item{Publicly available artifacts and a web service to enhance the development of a deep learning system\footnote{PROFET Service and Artifacts: \url{http://profet.ddps.cloud}}}
\end{itemize}

\begin{table}
    \def\arraystretch{1.1}
    \centering
        \begin{tabular}{ |c||c|c|c|c|  }
         \hline
         \multicolumn{5}{|c|}{AWS GPU Instance Specification} \\
         \hline
         Instance Family & G3s & G4(dn) & P2 & P3\\
         \hline
         \hline
         GPU Model & M60 & T4 & K80 & V100\\
         GPU Core & 2048 & 2560 & 2496 & 5120 \\
         GPU Clock(MHz) & 1178 & 1590 & 875 & 1380\\
         TFLOPS(FP32) & 4.825 & 8.141 & 4.113 & 14.13\\
         \hline
         Released Year & 2017 & 2019 & 2016 & 2017 \\
         Price(\$/hr) & 0.75 & 0.526 & 0.9 & 3.06\\
         \hline
        \end{tabular}
    \caption{Specification of different GPU Instances on AWS}
    \label{tab:aws_gpu_spec}
\end{table}

\section{Training CNN on Cloud}
Training a model can take weeks or months, depending on the internal architecture of the CNN implementation~\cite{dnn-workload-sensetime-short, dnn-workload-microsoft-short, alibaba-dl-train-workload-short}. Various configurations influence CNN model training time, and it is critical to understand the performance diversity of CNN training to build an optimal training environment.

\subsection{Overview of CNN}\label{sec:cnn-overview}
CNN algorithms are commonly used to analyze visual representations from a given input dataset, which typically consists of images or videos. The most common type of CNN model is built with input and output layers connected by a series of hidden layers. Each hidden layer includes a wide range of operations, such as convolution, activation, and pooling. The operations extract information from the previous layer and abstract it by applying a specific operation to pass it to the next layer. In the early days of CNN implementation, small models were proposed, such as LeNet5~\cite{lenet-short}. Recently, diverse and complex models regarding depths, layers, activation functions are proposed which improve the model accuracy drastically. They include AlexNet~\cite{alexnet-short}, VGG~\cite{vgg}, ResNet~\cite{resnet-short}, and Inception~\cite{inception}.

\begin{figure}[t]
    \centering
    \includegraphics[width=0.4\textwidth]{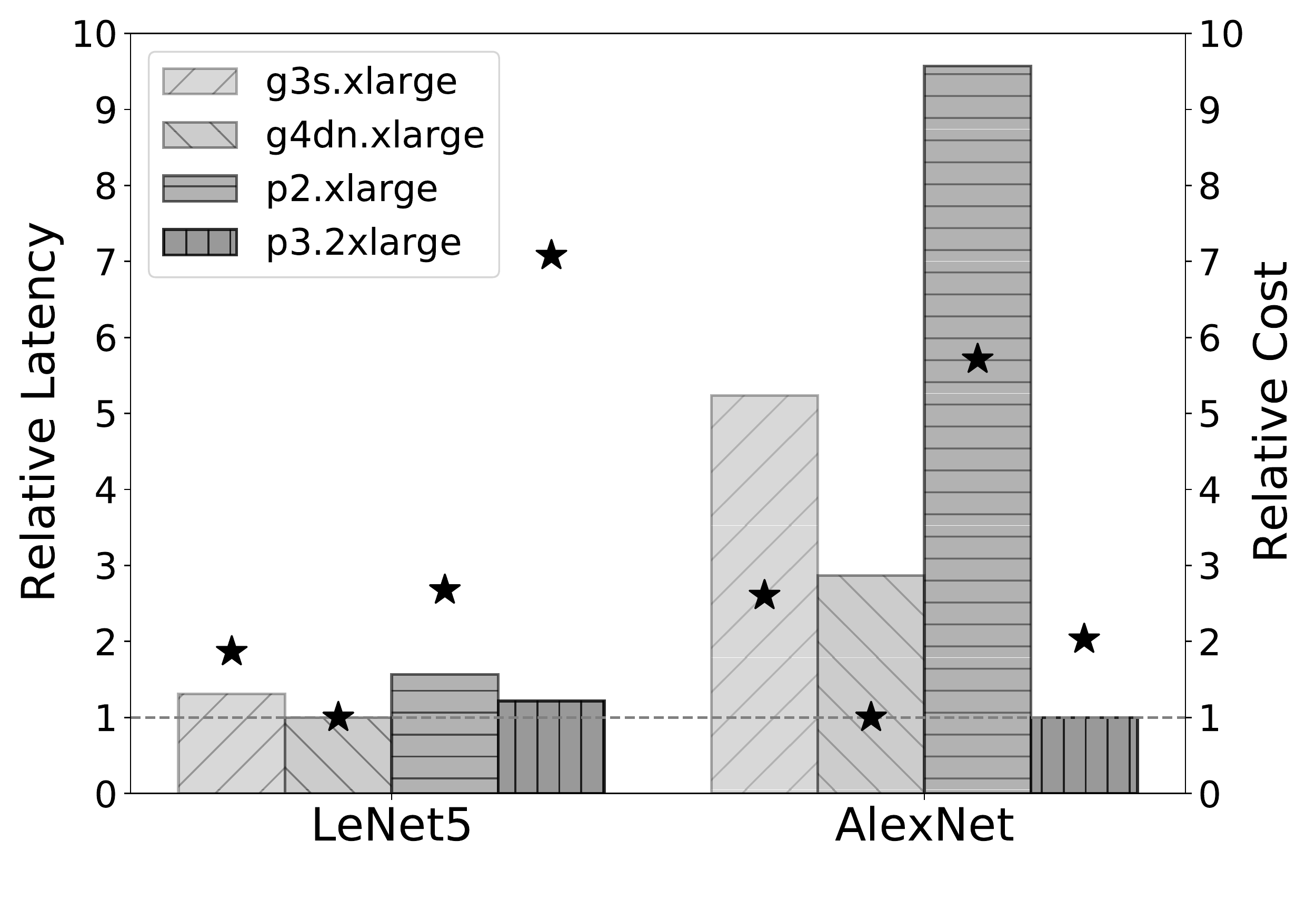}
    \caption{Performance variation of CNN model training using GPU cloud instances}
    \label{fig:motivation_model}
\end{figure}

\subsection{GPU Instances on Cloud}\label{sec:aws-gpu-instances}
As CNN model training task requires extensive computing capacity, GPU is widely used to train a model because of its superb performance and well-established Software Development Kit (SDK). Due to its high cost of hardware and cluster management overhead, using GPU devices through public cloud services becomes the norm, and most public cloud vendors provide the service. Similar to other cloud instances, there exist various instance types which equip GPU devices. Table~\ref{tab:aws_gpu_spec} shows GPU-based instance types provided by AWS. We show the smallest size in each instance family. As shown in the table, each instance type has its own GPU model manufactured by NVIDIA. CNN training performance is expected to be primarily determined by the processing capacity of GPU, which can be expressed in terms of the number of GPU cores, clock speed, and floating-point operations per second (FLOPS). Other than the distinct processing capacity of each instance type, the price also differs significantly, and it becomes challenging for CNN algorithm developers to decide optimal GPU cloud instances for their model development.

\subsection{Understanding CNN Training Performance}
The diversity of CNN model, configurations, and cloud instances can result in significant differences in model training time. To represent latency and cost difference under various settings, we conducted thorough experiments. Figure~\ref{fig:motivation_model} presents the mini-batch training time of \emph{LeNet5}~\cite{lenet-short} and \emph{AlexNet}~\cite{alexnet-short} on various GPU cloud instances presented in Table~\ref{tab:aws_gpu_spec}. The latencies are represented with bars in the primary vertical axis. The secondary vertical axis displays the relative cost of completing a given training workload, with the values represented by star marks. They are normalized to the least value in each workload. For each workload, the cloud instances are shown in the order of \emph{g3s}, \emph{g4dn}, \emph{p2}, and \emph{p3}. For \emph{LeNet5}, the \emph{g4dn} is the fastest, while the \emph{p3} is the fastest for \emph{AlexNet} training. Comparing the best and worst-performing instance types, the latency of \emph{LeNet5} is less than two times (\emph{g4dn} and \emph{p2}), but that of \emph{AlexNet} is close to ten times (\emph{p3} and \emph{p2}). Regarding the cost, \emph{g4dn} incurs the least cost for both workloads, but choosing \emph{p3} instance can be more beneficial for \emph{AlexNet} because its training latency is about one-third with only twice more cost than \emph{g4dn}.

Figure~\ref{fig:motivation_model} presents drastic CNN training latency variations on diverse cloud instances with GPU devices under different model architectures. Considering the continuous release of new cloud services, resources, and pricing mechanisms~\cite{deepspotcloud-short, spot-price-prediction-short, ri-marketplace-short}, it is challenging for CNN algorithm developers to follow new updates and timely apply them to their training environment. Furthermore, it is very cumbersome to try every instance type that might need custom device library setup to check how a CNN model performs. Therefore, a guidance of estimated performance of custom CNN algorithm implementations on diverse cloud instance types is mandatory to help developers to build an optimal training environment and focus on core application development.

\section{PROFET : Modeling Training Time}\label{section:profet-overall}
The goal of PROFET is to predict the training time for arbitrary CNN implementations on various GPU instances provided by a public cloud service vendor with minimal exposure to implementation details. Minimally exposing the implementation details is important, especially in a public cloud environment. Setting up a CNN development environment using GPUs requires considerable system operation effort, which can be quite challenging for an algorithm developer. Thus, the responsibility of maintaining and operating an optimal development environment should be imposed on the cloud service vendor, which agrees with the recent cloud computing evolution direction~\cite{cloud-evolution-short} represented by the serverless computing~\cite{serverless-backward-forward-short, function-bench-short}. To allow public cloud vendors to provide an optimal training environment for arbitrary CNN implementations, CNN model characteristics should be provided while hiding the detailed internal architecture, and PROFET achieves this goal using abstracted operation information.

To build a prediction model, a set of feature vectors which we denote as $\boldsymbol{X}$ is generated from offline experiments. $\boldsymbol{X}$ is composed of $N$ workloads. Each workload return a feature vector with a dimension of $D$. Thus, the dimension of $\boldsymbol{X}$ is $N \times D$. We denote each workload scenario as $\boldsymbol{x}_i, i=1:N$. To note $j$-th feature of workload $i$, where $i=1:N, j=1:D$, we use $x_{ij}$. 

To generate diverse CNN workloads, we variate CNN training scenarios with respect to the instance types ($G$), model architectures ($M$), batch sizes ($B$), and input image pixel sizes ($P$). For $G$, we assume GPU cloud instances provided by AWS, $G \owns \{g3s, g4dn, p2, p3\}$. For models, $M$, we used well-known CNN models in literature, $M \owns$ \{\emph{AlexNet}, \emph{LeNet5}, \emph{InceptionV3}, \emph{InceptionResNetV2}, \emph{MobileNetV2}, \emph{MNIST\_CNN}, \emph{CIFAR10\_CNN}, \emph{ResNetSmall}, \emph{ResNet18}, \emph{ResNet34}, \emph{ResNet50}, \emph{VGG11}, \emph{VGG13}, \emph{VGG16}, \emph{VGG19}\}. We used five batch sizes, $B \owns \{16, 32, 64, 128, 256\}$, and five input image pixel sizes, $P \owns \{32 \times 32,  64\times 64, 128 \times 128, 224 \times 224, 256 \times 256 \}$. To generate images with different pixel sizes, we used the Numpy library. All workload scenarios can be generated by conducting Cartesian product in the four dimensions ($G \times M \times B \times P = \{(g, m, b, p) : g \in G, m \in M, b \in B, p \in P\}$). All cases of $G \times M \times B \times P$ cannot be completed due to hardware or model constraints. Filtering out inexecutable cases, we finalize $1228$ cases which becomes the cardinality of input dataset, which is $N$. Each workload $\boldsymbol{x_i}$ is a vector with a length $D$ that is returned by the underlying profiler. Among the operations provided by TensorFlow Profiler~\cite{tf-profiler}, we generate $65$ aggregated high-level operations from $1228$ executable workloads. Thus, the dimension of $\boldsymbol{X}$ is $1228 \times 65$.

\subsection{Extracting CNN Characteristics on Cloud}\label{ch:feature-profiling}
A PaaS cloud model frees CNN algorithm developers from operating cloud infrastructures and DNN development platforms because a cloud service provider is responsible for maintaining them. To provide efficient deep learning platforms, cloud vendors should understand distinct workload characteristics submitted by clients. However, in a PaaS model, most CNN algorithm developers are not willing to share source codes or internal architectures of a developed model. Thus, previous works which reference internal model architectures to predict DNN training latency~\cite{paleo-short, neuralpower-short, mlpredict-bigdata-short} cannot be applied directly on a PaaS environment. To predict training latency of arbitrary CNN implementations on a public PaaS model, PROFET should be able to characterize a workload while minimally disclosing the internal model architecture. 

\begin{figure}[t]
    \centering
    \includegraphics[width=0.45\textwidth]{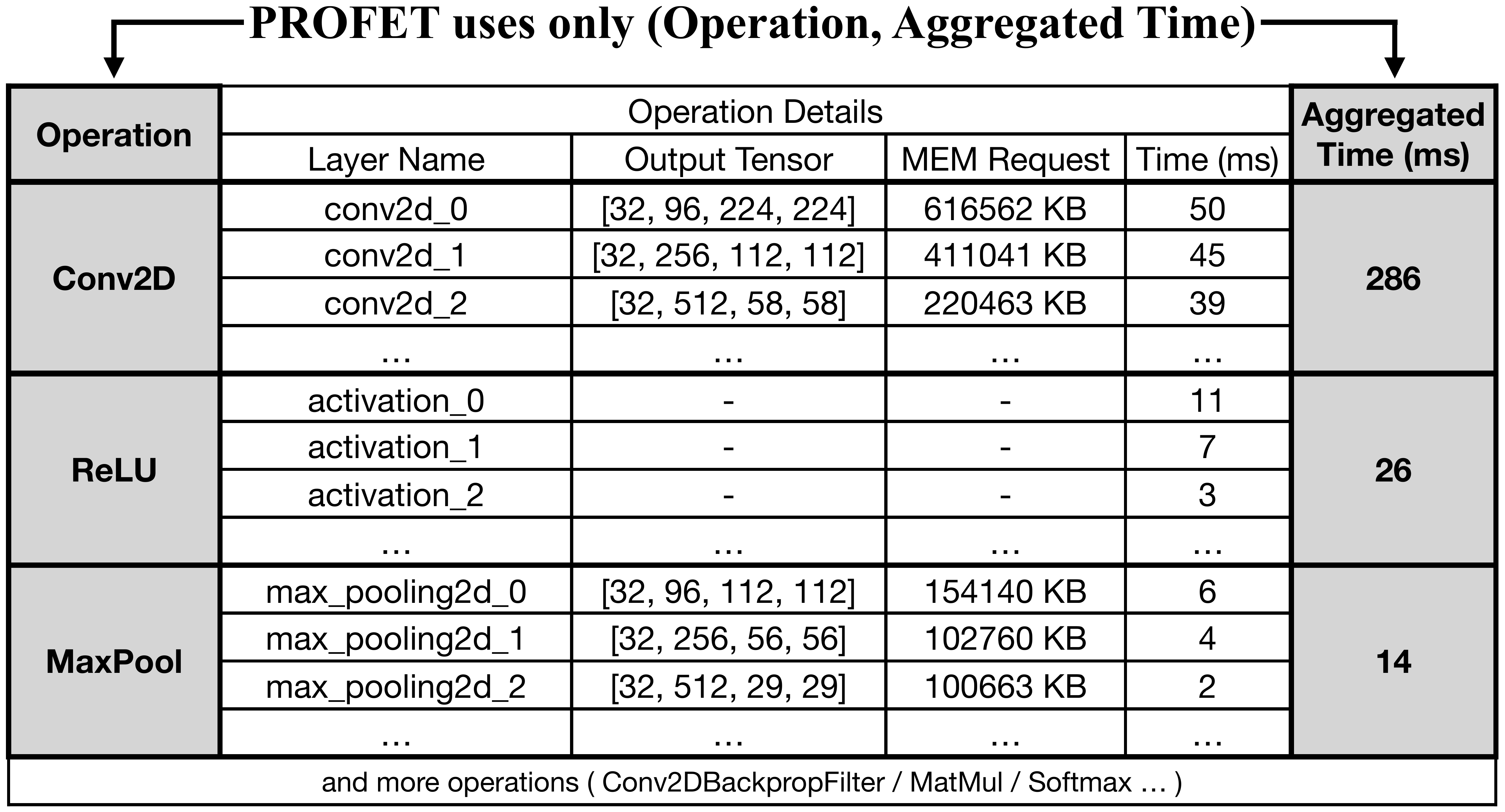}
    \caption{An example of profiling data generated from AlextNet model training using TensorFlow profiler}
    \label{fig:tf-profiler-example}
\end{figure}

Other than referencing source code to characterize CNN workloads, PROFET proposes to use performance metrics generated during a training phase provided by a profiler of DNN programming platforms, such as TensorFlow~\cite{tensorflow-short}, MXNet~\cite{mxnet-short}, Torch~\cite{pytorch-short}, and Theano~\cite{theano-short}. Though information from metrics of each platform differs slightly, they provide response times for core DNN-specific operations in common.

To better understand performance metrics provided by a DNN development platform, we show an example of a CNN model profiling outcome generated by TensorFlow Profiler~\cite{tf-profiler} in Figure~\ref{fig:tf-profiler-example}. A profiling output contains Operation, Operation detail, and Latency fields. The operation field indicates a method name used in a source code which is specified by a development platform. The operation details field contains rich information of a CNN model which includes method name and its layer location, input and output tensor sizes, memory usage, and many more. The information contains internal architecture of custom implementation, and a model can be assembled using the information. Of the profiling outcome, PROFET proposes to use the operation field and the corresponding aggregated time field as features to represent CNN workloads and make a model to predict training time of arbitrary CNN implementations. With the high level information about operations, Hafeez et. al. proved that profiled outcome can well represent characteristics of arbitrary CNN models~\cite{aws-cnn-analysis-short}. Furthermore, hiding operation details can relieve CNN algorithm developers when they share the workload characters with public cloud PaaS vendors which can offer an optimal development environment for any kinds of CNN workloads. In summary, PROFET adopts a \emph{black-box} approach by using high-level expression of CNN workloads without using internal model architectures which is contrary to the \emph{white-box} approach adopted in previous work~\cite{paleo-short, mlpredict-bigdata-short, neuralpower-short}, and such black-box design fits very well with the public cloud environment where the service providers and consumers differ.

\begin{figure*}
    \centering
    \includegraphics[width=0.95\textwidth]{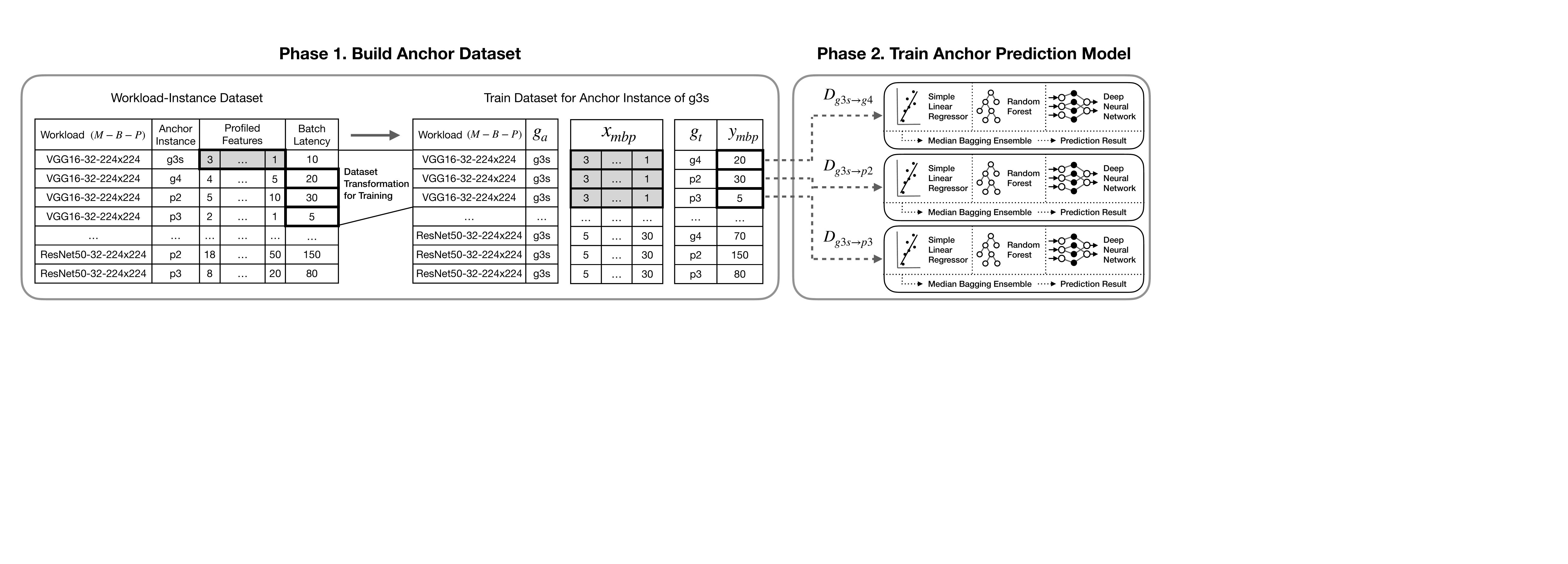}
    \caption{Predicting the training latency on a target instance type based on profiled feature from an anchor instance}
    \label{fig:anchor_prediction}
\end{figure*}

Using the operation names and aggregated latencies as features constructs input vectors, $\boldsymbol{X}$, and we define output vectors, $\boldsymbol{Y}$, as a corresponding batch latencies of the input vectors. Formally speaking, for an arbitrary CNN workload profiling features, $x_i$, $y_i$ is a scalar value which represents a measured batch latency of workload $i$ where $i=1:N$. To note a value of specific feature, $j$, of model $i$, we use $x_{ij}$, where $i=1:N, j=1:D$, which is a scalar value. Creating feature vectors, $\boldsymbol{X}$, from a deep learning platform incur non-negligible extra overhead that can impact batch latency, $\boldsymbol{Y}$. In our off-line experiments, about $20\% - 30\%$ larger batch latency is measured when we enabled profiling. To remove the impact from the profiling overhead, we conducted two sets of experiments with a workload, $i$, one with profiling enabled, another without profiling enabled. In the experiments with profiling, we gather $\boldsymbol{X}$. To get an accurate value of $\boldsymbol{Y}$, we measured the batch latency without enabling profiling. This procedure of separately generating $\boldsymbol{X}$ and $\boldsymbol{Y}$ is identical when a new CNN workload scenario is predicted with PROFET; an user enables profiling to get a feature vector to get predicted batch training latency without profiling enabled.

\subsection{Modeling CNN Performance on Cloud}\label{ch:cross-instance-modeling}
PROFET predicts the training latency of arbitrary CNN implementations ($M$) on different GPU-based instance types ($G$), batch sizes ($B$), and input image pixel sizes ($P$). To formally express input datasets, $x_i$, in a finer-grained way, we use the lowercase letter of each category as a subscript, $x_{mgbp} = \{(m, g, b, p) : m \in M, g \in G, b \in B, p \in P\}$. For example, to specify input datasets of an arbitrary model ($m$), batch size ($b$), and input pixel size ($p$) for all the possible GPU instances in $G$, we use $x_{mbp}$ which excludes $g$ from the subscript, and the cardinality of $x_{mbp}$ equals the size of $\arrowvert G \arrowvert$, which means four instance types with GPU devices used in this proposed work.

The cross-instance performance model predicts the training latency of an arbitrary workload of $x_{mbp}$ on various instance types. This model requires a profiled feature set of a workload, $x_{mbp}$, on an instance type which we call an anchor instance, $g_a$. Using the profiling outcome from $g_a$, PROFET predicts the training latency on a target instance type, $g_t$ where $g_a \neq g_t$ and $g_a,g_t \in G$, of workload $x_{mbp}$. Let us define a training latency prediction model from $g_a$ to $g_t$, as $f_{g_a \rightarrow g_t}$. The training dataset of the function is defined as $\mathcal{D}_{g_a \rightarrow g_t}$ which is defined as follows.

\begin{equation*}
\mathcal{D}_{g_a \rightarrow g_t} \triangleq \{(x_{mbp|G=g_a},y_{mbp|G=g_t}) | mbp \in (M \cup B \cup P)\}
\end{equation*}

For an arbitrary workload of $mbp$, an input feature vector $x$ is returned from a profiler after executing on an anchor instance, $g_a$. The output of the model, $y$, is a scalar value that is a batch latency for the same workload, $mbp$, executed on a target instance, $g_t$.

With training dataset, $\mathcal{D}_{g_a \rightarrow g_t}$, PROFET builds a prediction model using an ensemble algorithm~\cite{bagging} with a median operator~\cite{median-ensemble}. In machine learning, ensemble modeling combines multiple individual weak models to have higher prediction accuracy. The bootstrap aggregating (bagging)~\cite{bagging} trains multiple models using the same set of train dataset and allocate weights for each model to come up with the final prediction. Lang et. al.~\cite{median-ensemble} proposed a median-based bagging algorithm that adopts the median predicted values among multiple predictive values. According to the authors, using the median value improves model accuracy by removing noise in real-world signal processing applications. For ensembling, PROFET builds three independent prediction models of a DNN, random forest~\cite{random-forest}, and linear regressor. We build a DNN model of $128 \times 64 \times 32 \times 16 \times 1$ dense layer architecture with ReLU activation~\cite{relu-short} in each layer while minimizing the combined loss of Mean Absolute Percentage Error (MAPE) and Root Mean Squared Error (RMSE) with the Adam optimizer~\cite{adam-optimizer}. To build random forest and linear regression models, we use Python's Scikit-Learn library~\cite{scikit-learn} with the default hyper-parameters provided by the library.

Figure~\ref{fig:anchor_prediction} explains the overall procedure of cross-instance batch latency prediction. The left side of the figure(phase 1) expresses the generation of the training dataset. In the step, all the workloads are executed on all the GPU instance types in $G$, and profiling features with the corresponding batch latencies are recorded. To use the experimental result as a model training input, we match the profiled feature ($x_{mbp}$) of an anchor instance ($x_g$) and batch latencies ($y_{mbp}$) of target instances ($x_t$). In the figure, the anchor instance type is \emph{g3s}, and the target instance types are \emph{g4, p2, p3}. It shows three training input cases; for a same profiled feature from \emph{g3}, three batch latencies from \emph{g4, p2, p3} are matched as outputs. In the prediction model building step, separate models are trained per each anchor and target instance type combination using distinct training dataset of $\mathcal{D}_{g_a \rightarrow g_t}$. Using the unique dataset per anchor and target instance types, three models of linear regressor, random forest, DNN are trained for median-ensemble modeling. In the figure, three separate ensemble models are built, $f_{g3s \rightarrow g4}, f_{g3s \rightarrow p2}$, and $f_{g3s \rightarrow p3}$.

\subsubsection{Predicting Latency on New GPU Types}\label{sec:new-gpu-support}
A new type of GPU is released quite often, and it is important to anticipate training latency on such new hardware which was not available during the prediction modeling. PROFET does not use hardware specifications as prediction model features as previous work did~\cite{paleo-short, mlpredict-bigdata-short, habitat-short}, and it cannot predict training latency for a new GPU device which was not available at a modeling phase. However, this does not become a significant issue when the prediction service is provided by a cloud vendor. In the cloud, service providers decide when to make a new instance type available to customers, and they have enough time to prepare a latency prediction model for the new hardware. As cloud computing becomes more prevalent for diverse applications, cloud vendors should become more responsible for dealing with complex system operations. From this context, preparation of prediction model by a cloud service vendor for a new hardware type is rational.

\begin{figure*}[t]
    \centering
    \subfloat[g3s.xlarge]{
        \includegraphics[width=0.23\textwidth]{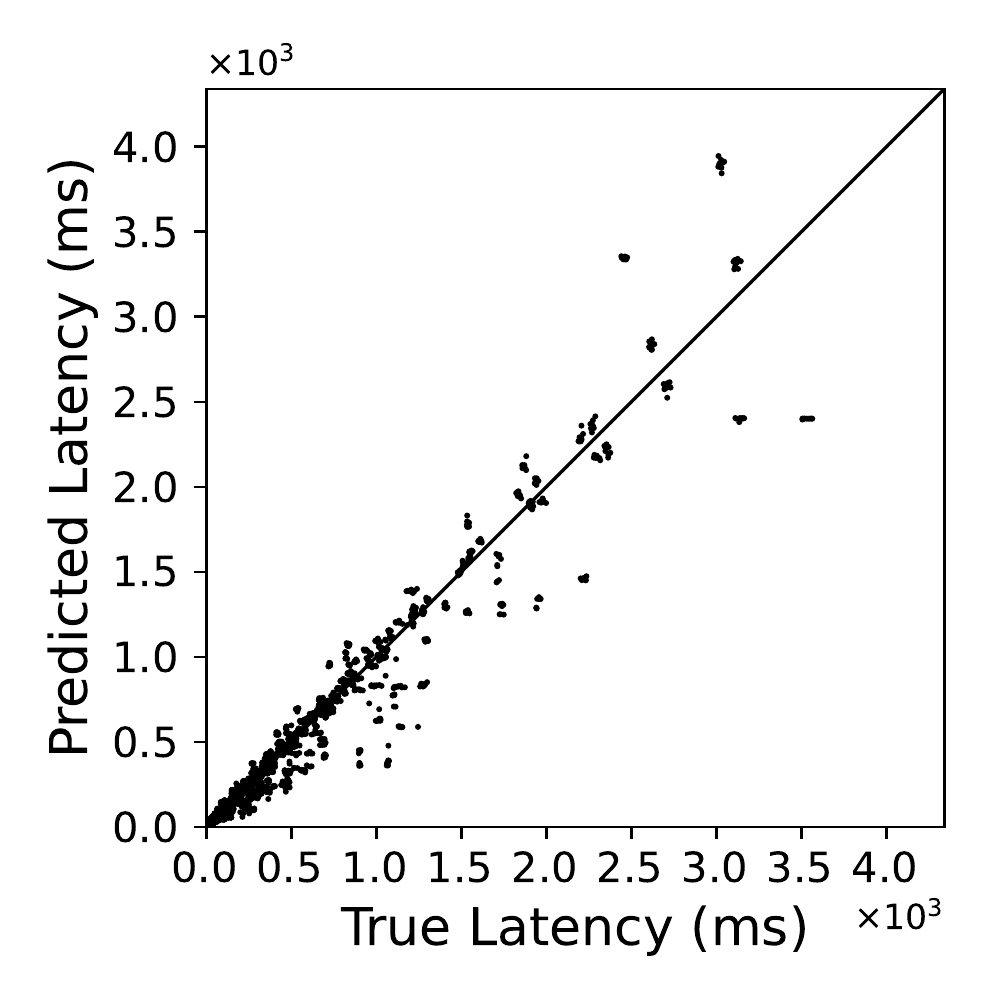}
        \label{fig:anchor-eval-g3s}
    }
    \subfloat[g4.xlarge]{
        \includegraphics[width=0.23\textwidth]{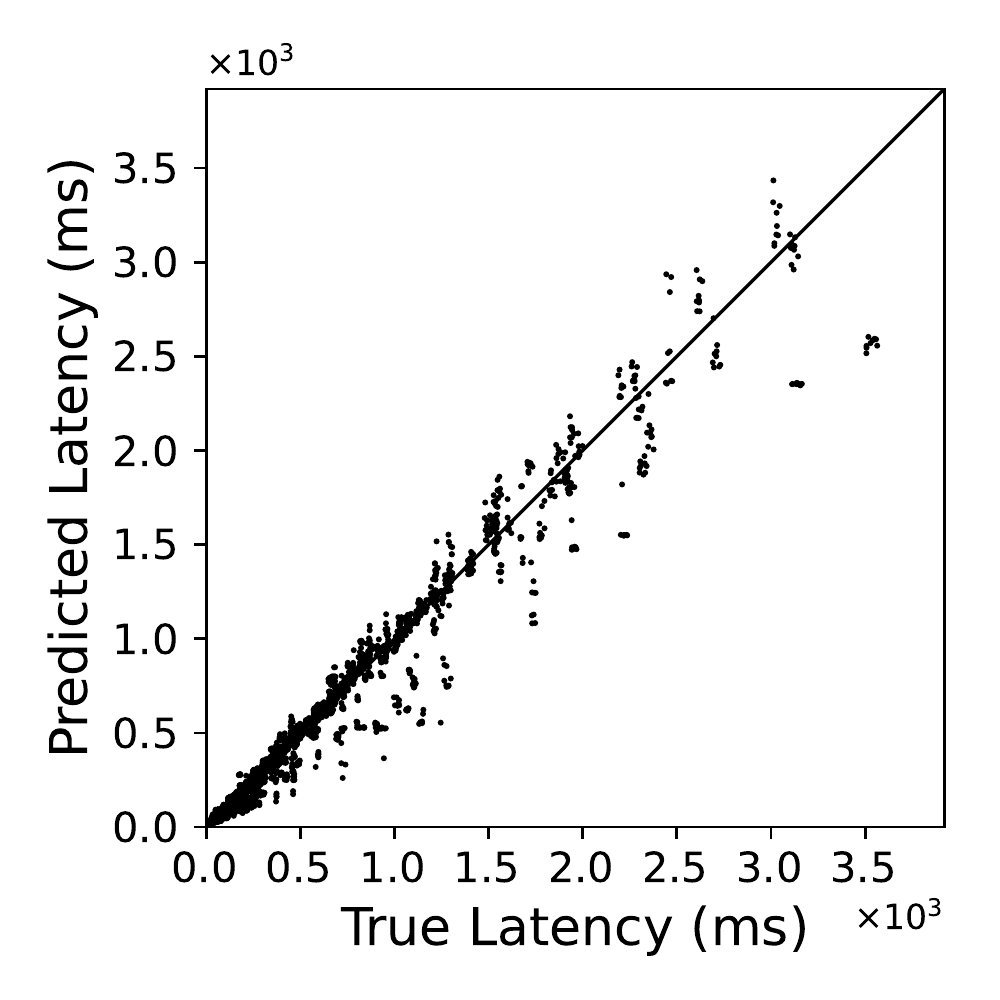}
        \label{fig:anchor-eval-g4}
    }
    \subfloat[p2.xlarge]{
        \includegraphics[width=0.23\textwidth]{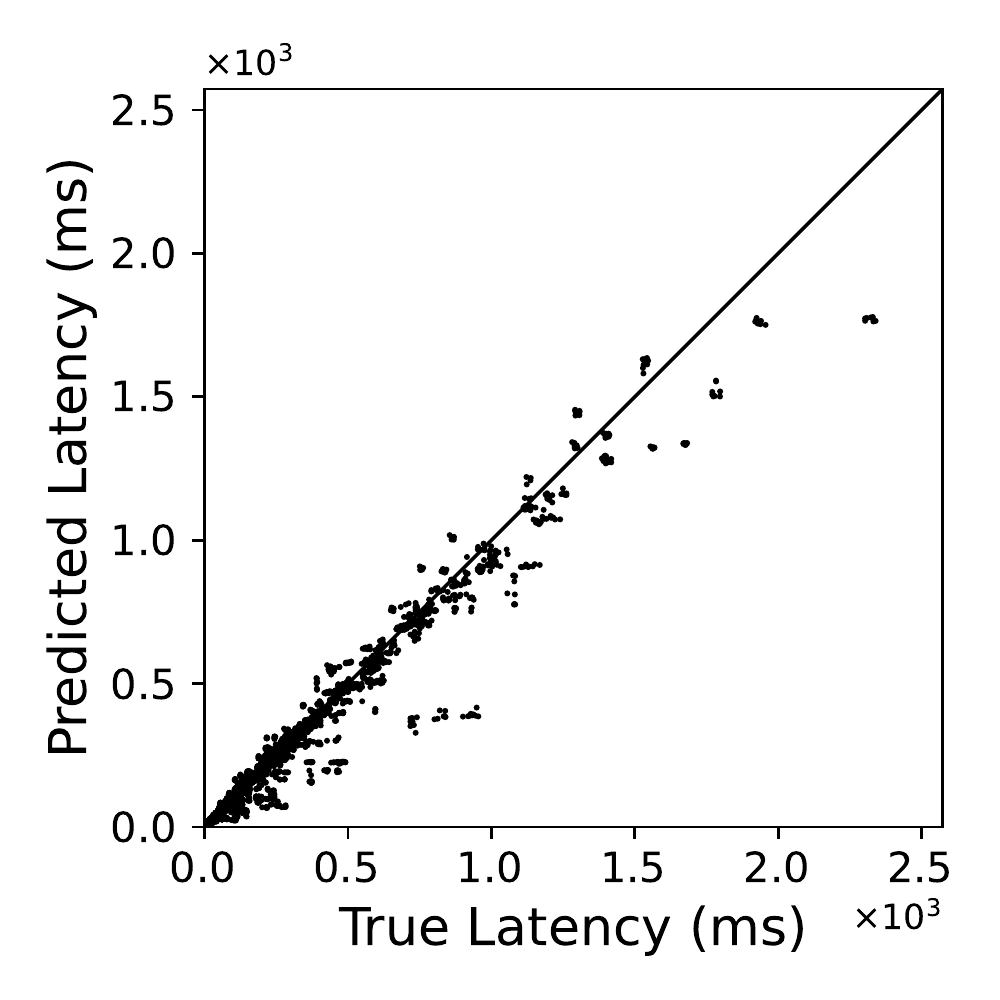}
        \label{fig:anchor-eval-p2}
    }
    \subfloat[p3.xlarge]{
        \includegraphics[width=0.23\textwidth]{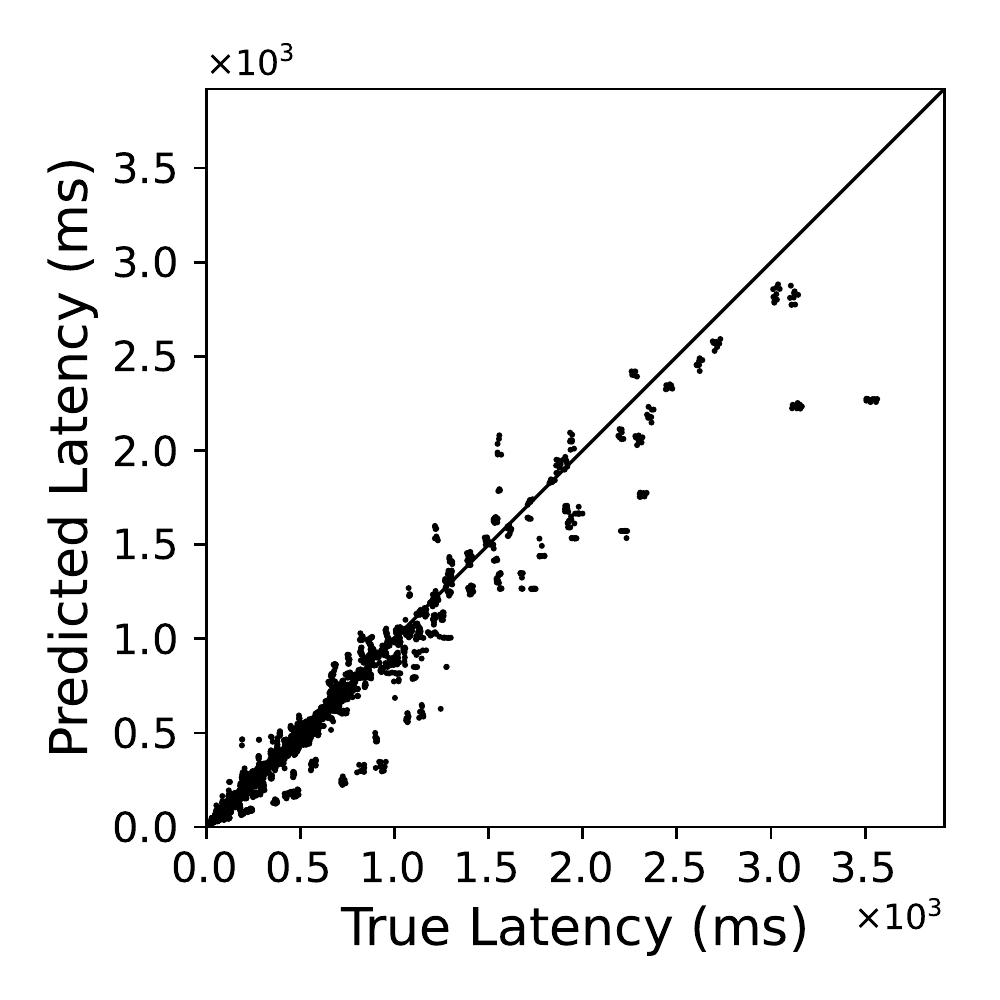}
        \label{fig:anchor-eval-p3}
    }
    \caption{True (x-axis) and predicted (y-axis) latencies for different anchor instances}
    \label{fig:anchor-evaluation}
\end{figure*}

\section{Evaluation}
To evaluate the accuracy of PROFET, we mainly use MAPE, RMSE, and coefficient of determination which is denoted as $R^2$. The $R^2$ metric represents a quantitative measurement of how the predicted outcome from a regressor model resembles the true value, and the higher $R^2$ means better accuracy.

The experiments are conducted on AWS GPU instances with various CNN models, batch sizes, and input image pixel sizes, which are defined in Section~\ref{section:profet-overall}. In running CNN training on AWS, we used Deep Learning AMI (Ubuntu 18.04) Version $35.0$ which contains NVIDIA GPU driver (450.80.02), CUDA SDK (10.1), and TensorFlow (2.3.0). According to the different evaluation criteria, the models to be predicted are completely removed from the training datasets.

\subsection{Performance of Cross-Instance Latency Prediction}
Figure~\ref{fig:anchor-evaluation} presents the prediction accuracy of cross-instance performance modeling presented in Chapter~\ref{ch:cross-instance-modeling}. Each sub-figure represent different anchor instance; Figures~\ref{fig:anchor-eval-g3s},~\ref{fig:anchor-eval-g4},~\ref{fig:anchor-eval-p2},~\ref{fig:anchor-eval-p3} represents the anchor instances ($g_a$) of \emph{g3s.xlarge, g4dn.xalrge, p2.xlarge}, and \emph{p3.2xlarge}, respectively. The horizontal axis in the figure represents the true training latency, while the vertical axis represents the predicted latency by PROFET. A scatter plot is used to express the real and predicted values, and a value close to an equation $y=x$ indicates an accurate prediction. As shown in the figure, the predicted values are close to the real CNN training time which presents superb prediction accuracy of PROFET.

\begin{figure}
    \centering
    \includegraphics[width=0.4\textwidth]{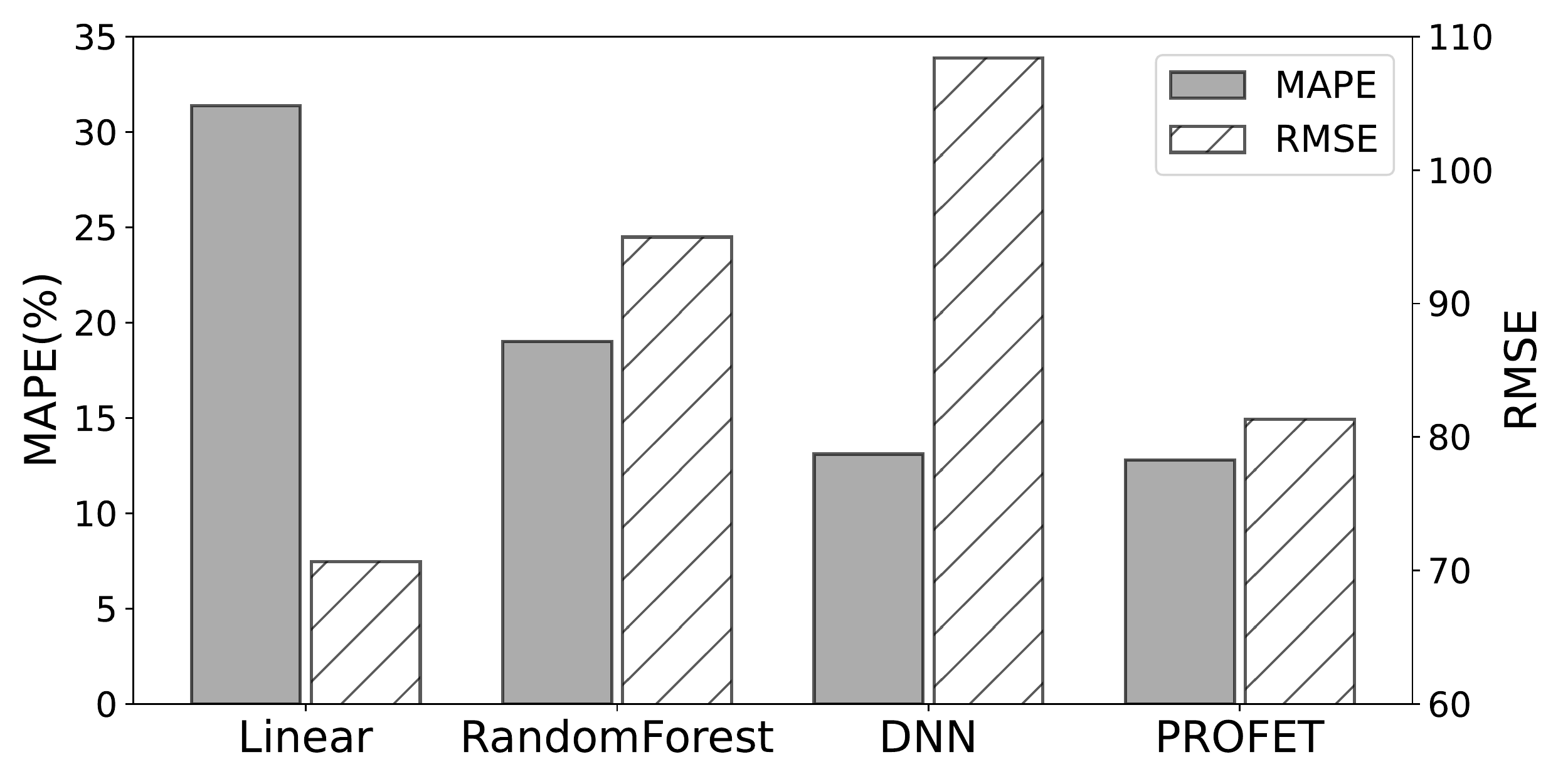}
    \caption{Higher prediction accuracy of the proposed median ensemble approach of PROFET}
    \label{fig:performance-model-efficiency}
\end{figure}

To quantitatively evaluate the prediction accuracy of the median ensemble modeling of PROFET, we compare the MAPE, and RMSE with other approaches in Figure~\ref{fig:performance-model-efficiency}. The primary vertical axis represents the MAPE whose values are shown in solid gray bars. The secondary vertical axis shows the RMSE whose values are shown in the bars with the right upper diagonal pattern. For both MAPE and RMSE, lower values are better. In the horizontal axis, we show distinct prediction models of a linear regressor (\emph{Linear}), a tree-based non-linear regressor (\emph{RandomForest}), \emph{DNN}, and the proposed ensemble algorithm (\emph{PROFET}). In the \emph{Linear} approach, different from others, we use batch latency measured from the anchor instance as input data. The linear prediction model is an order-1 regressor which can be expressed as $\alpha x + \beta$, where $\alpha$ means coefficient of batch latency, and $\beta$ means a bias value of the model. Details of \emph{RandomForest} and \emph{DNN} implementations are presented in Section~\ref{ch:cross-instance-modeling}. Using the three different models, the median ensemble approach of PROFET takes the three predicted values and determines the median value as the predicted latency. 

As shown in the figure, PROFET shows the lowest error in MAPE which is a $12.8\%$. Comparing to the MAPE of \emph{DNN}, the lowest MAPE among single models, the PROFET shows a $2.4\%$ improvement. But in the case of RMSE, PROFET shows a $24.56\%$ better prediction result than a single \emph{DNN} model. Using three distinct models with different complexity in an ensemble manner compensates for the weakness of each model, and it greatly improves the generality of the prediction model. For instance, the MAPE and RMSE of \emph{Linear} model show drastic difference. Careful investigation reveals that the \emph{Linear} model shows poor prediction accuracy for small models which can make MAPE worse even for small error values.

\subsection{Comparison to the State-of-the-Art}
Most contemporary CNN model training latency prediction algorithms adopt the \emph{white-box} approach while referencing the internal model architecture and hardware configurations as features. We qualitatively argue that the \emph{black-box} approach of the PROFET is more appropriate in a public cloud environment. To quantitatively present the prediction accuracy of PROFET with the most recent related work, we compare its performance with Paleo~\cite{paleo-short}, MLPredict~\cite{mlpredict-bigdata-short}, and Habitat~\cite{habitat-short}.

First, reproducing the Paleo experiment result necessitates the creation of a training environment that is identical to the one used by the Paleo authors. We discovered that building an identical environment is impractical due to incompatible TensorFlow, GPU driver and library versions. Applying Paleo's prediction algorithm on a contemporary development environment does not result in the as accurate result as one presented in the original paper, and we decided to compare the result presented in the original publication~\cite{paleo-short}. Table~\ref{table:paleo-profet-compare} compare the prediction accuracy of PROFET and Paleo for common models which are AlexNet~\cite{alexnet-short} and VGG-16~\cite{vgg}. As shown in the table, PROFET outperforms Paleo. For MAPE, PROFET is $67.85\%$ better and $45.78\%$ better for RMSE.

\begin{table}
    \def\arraystretch{1.35}
    \centering
    \begin{tabularx}{0.45\textwidth}{|c||X|X|}
    \hline
              & \makecell{PALEO}   & \makecell{PROFET}  \\ \hline
        MAPE  & \makecell{10.11} & \makecell{3.25} \\
        $R^2$ & \makecell{0.99} & \makecell{0.99} \\
        RMSE  & \makecell{32.36} & \makecell{17.55} \\ \hline
    \end{tabularx}
    \caption{The prediction accuracy of Paleo and PROFET}
    \label{table:paleo-profet-compare}
\end{table}

MLPredict~\cite{mlpredict-bigdata-short} presented an algorithm to predict training time of DNN models across different GPU devices for diverse batch sizes and showed better performance. To compare the prediction accuracy of PROFET with MLPredict, we implemeted the MLPredict algorithm following the paper. Table~\ref{tab:mlpred-profet-compare} shows the MAPE and RMSE of MLPredict and PROFET with different batch sizes with \emph{VGG16} which showed the best performance in~\cite{mlpredict-bigdata-short}. As shown in the table~\ref{tab:mlpred-profet-compare}, the prediction accuracy of PROFET outperforms MLPredict for both MAPE and RMSE. We could observe that the accuracy of MLPredict is poorer than the result presented in the paper~\cite{mlpredict-bigdata-short}. In the original MLPredict paper, the authors mainly predicted training latency of small batch sizes, in the range of $1$ to $16$. The error rate of MLPredict increases as the batch size becomes larger, and we could conclude the MLPredict algorithm is rather optimized for small batch sizes. However, such small batch sizes are impractical in real-world model training because they increase training time. Recent GPUs available device memory can accommodate larger batch sizes for many well-known DNN implementations, and we believe it is more important to accurately predict the training time of larger batch sizes. In summary, PROFET improves the RMSE by $81.54\%$ comparing to MLPredict. 

Last, we compare PROFET with the most recent related work, Habitat~\cite{habitat-short}, which uses detailed profiling results to build a model and predict the training latency across different GPUs. Habitat does not support prediction for varying batch sizes, and we measured the prediction accuracy with fixed batch sizes of $16, 32$, and $64$ both for PROFET and Habitat. Habitat's system implementation is open-sourced, and we could reproduce the experimental results in the publication. Table~\ref{table:habitat-profet-compare} presents the prediction accuracy of PROFET and Habitat. For both systems, we use two GPUs of Nvidia T4 and V100. The row with T4 $\rightarrow$ V100 indicates that the anchor instance is T4, and the target instance to predict the training latency is V100. For different anchors and target GPUs, we predict the training latency of \emph{Resnet50, InceptionV3}, and \emph{VGG16} models and present the average MAPE. We select the GPU device and CNN models that are common to the PROFET and Habitat experiments. On average MAPE, PROFET shows $32.26\%$ lower than that of Habitat.

In summary, PROFET presents the best prediction accuracy among contemporary related work by lowering the MAPE by $32\%$ (Habitat), $68\%$ (Paleo), and $82\%$ (MLPredict). We claim that the higher prediction accuracy of PROFET while using less detailed information than the \emph{white-box} approach is due to the creative definition of input features from the anchor instance and target latency from the instance type that PROFET needs to predict.

\begin{table}
    \def\arraystretch{1.1}
    \centering
        \begin{tabularx}{0.45\textwidth}{ |c||X|X|X|X|  }
         \hline
         & \multicolumn{2}{c|}{MAPE (\%)} & \multicolumn{2}{c|}{RMSE} \\
         \hline
         BS & MLPredict & PROFET & MLPredict & PROFET \\
         \hline
         16                   &\makecell{15.68}        &\makecell{2.96}   &\makecell{90.82}      &\makecell{8.78} \\
         32                   &\makecell{17.89}        &\makecell{3.25}   &\makecell{135.63}     &\makecell{17.55} \\
         64                   &\makecell{24.27}        &\makecell{4.45}   &\makecell{408.81}     &\makecell{70.51} \\
        \hline
        \end{tabularx}
    \caption{The prediction accuracy of VGG16 model for diverse GPU instances of MLPredict and PROFET}
    \label{tab:mlpred-profet-compare}
\end{table}

\begin{table}[t]
    \def\arraystretch{1.1}
    \centering
    \begin{tabularx}{0.45\textwidth}{|c||X|X|}
    \hline
                                & \makecell{Habitat}    & \makecell{PROFET} \\ \hline
        T4 $\rightarrow$ V100   & \makecell{12.16}      & \makecell{4.07}   \\
        V100 $\rightarrow$ T4   & \makecell{7.99}       & \makecell{8.15}   \\ \hline
    \end{tabularx}
    \caption{The prediction accuracy (MAPE) of Habitat and PROFET with different combination of anchor-target GPUs}
    \label{table:habitat-profet-compare}
\end{table}

\section{Related Work}
\textbf{Performance Modeling on Various Hardware}: Paleo~\cite{paleo-short}, MLPredict~\cite{mlpredict-bigdata-short} and NeuralPower~\cite{neuralpower-short} proposed performance model for DNN training job. To accurately predict the training time or power usage, they use the internal model architecture and GPU specification as input feature of prediction model. Most recent work, Habitat~\cite{habitat-short} uses a profiling result as PROFET does. However, Habitat uses detailed profiling output which might be inappropriate in a public cloud environment. The aforementioned works refer to the internal architecture of target models (\emph{white-box}), and algorithm developers may be hesitant to share the architecture. 

\textbf{Using Profiler for DNN Performance Analysis}: TBD~\cite{tbd-short} and Yeung et al.~\cite{gpu-util-cloud-short} examined the profiling results of various DNN models, hardware, and frameworks and presented an analysis of processing throughput, utilization, and memory consumption of each workload. PerfNetV2~\cite{perfnet-v2} and Ceer~\cite{aws-cnn-analysis-short} uses TensorFlow profiler to collect detailed DNN operation data to build a performance model, and it is similar to PROFET. But it is hard to compare performance with Ceer due to the lack of publicly available artifacts.

\textbf{DNN System with Performance Modeling}: Optimus~\cite{optimus}, Cynthia~\cite{cynthia}, RubberBand~\cite{rubberband}, and Chaudhary et. al.~\cite{dl-gpu-cluster-sharing} proposed a DNN tuning and training system that maximizes DNN cluster utilization to reduce processing time. They use the performance model to allocate DNN workload efficiently, or reduce the scale of the GPU cluster. Accurate performance estimation is critical for completing the aforementioned work, and PROFET is complementary to the work in that it can provide the accurate training time of various CNN implemetations.

\section{Discussions}
\textbf{Modeling train latency on diverse development platform}: During PROFET evaluation, we discovered that applying previous work algorithms on the most recent development platforms did not reproduce the prediction accuracy shown in the original papers, so we had to create a development environment using specific older versions. Most of previous work mentioned the version of development platforms for their evaluations, such as TensorFlow, Keras, GPU driver, and CUDA SDK, but we suffered from replaying the environments. We anticipate a similar situation for PROFET. There is not yet a system that can model the training latency as developers update underlying SDK libraries. Though it can be challenging, we believe that modeling performance with respect to a different SDK version can be useful, particularly for algorithm developers who need to decide whether an SDK version upgrade is worth the effort.

\textbf{Training latency prediction when using multiple GPU devices}: A large portion of DNN training tasks is conducted with single GPU. According to Microsoft Philly trace~\cite{dnn-workload-microsoft-short} and SenseTime trace~\cite{dnn-workload-sensetime-short}, $86\%$ and $70\%$ of DNN tasks are conducted on a single GPU, respectively. Though the majority of DNN training tasks are executed on a single GPU, using multiple GPUs can be an option to expedite a training task. In literature, many work discussed predicting the performance of parallel DNN training from a single GPU execution result~\cite{aws-cnn-analysis-short, iter-pred-multigpu}. Marble~\cite{marble-multigpu-dnn-scheduler-hpc-short} and Kahira et. al.~\cite{oracle-parallel-dnn-training-short} proposed optimal scheduling algorithms for parallel DNN executions. The work revealed that the bandwidth of CPU-GPU and GPU-GPU is an important metric for parallel training performance, and we are sure that applying communication overhead to PROFET will result in accurate prediction for training with multiple GPUs.

\textbf{Training latency prediction for non-CNN models} : The current PROFET supports training latency prediction of CNN models. Different types of DNN models, such as language processing models of Transformer~\cite{transformer-short} and BERT~\cite{bert}, use different operations, and the latency prediction models built with CNN profiling results can not apply to non-CNN models directly. We are currently working on extending PROFET to support general DNN models.

\section{Conclusion}
Training a CNN model with a GPU device has become the norm because it requires significant computing power, and public cloud service vendors offer a variety of GPU devices elastically. Due to the dynamically changing performance of arbitrary CNN implementations on various GPU devices, it is difficult for an algorithm developer to create an optimal training environment. This paper presented PROFET, which can predict the training latency of arbitrary CNN implementations with diverse configurations on various GPU devices, to aid in the development of an efficient CNN training environment. Without revealing implementation details of CNN implementation, PROFET can predict training latency on multiple distinct GPU devices. Thorough experiments reveal PROFET outperforms contemporary related work. Other than the quantitative superiority of PROFET, predicting latency without revealing implementation detail makes PROFET suitable in a cloud where an algorithm developer and resource providers differ. 

\section*{Acknowledgment}
This work is supported by the National Research Foundation (NRF) Grant funded by the Korean Government (MSIP) (No. NRF-2022R1A5A7000765 and NRF-2020R1A2C1102544), AWS Cloud Credits for Research program, and the SW Star Lab (RS-2022-00144309) of IITP.

\bibliographystyle{unsrt}
\bibliography{dl-cloud-recommend}

\end{document}